\documentclass[letterpaper, 10 pt, conference]{ieeeconf}
\usepackage{mathtools, cuted}
\mathtoolsset{showonlyrefs}
\usepackage{amssymb}
\usepackage{balance}
\usepackage{mathrsfs}
\usepackage{comment}
\let\labelindent\relax
\usepackage{enumitem}
\usepackage{mathdots}
\usepackage{xfrac}
\usepackage{bm}
\usepackage{array}
\usepackage{color}
\usepackage{algorithm}
\usepackage{algorithmicx}
\usepackage{algpseudocode}
\usepackage{physics}
\usepackage{comment}
\usepackage{lipsum}
\usepackage{float}
\usepackage{bbm}

\newtheorem{theorem}{Theorem}

\newcommand{\prt}[1]{\ensuremath{\left(  #1 \right)  }}

\DeclareMathAlphabet\mathbfcal{OMS}{cmsy}{b}{n}

\newtheorem{assumption}{Assumption}
\usepackage{etoolbox}

\providecommand{\keywords}[1]{\textbf{\textit{Index terms---}}}

\IEEEoverridecommandlockouts                              
\usepackage{eso-pic}
\AddToShipoutPictureBG*{%
	\AtPageUpperLeft{%
		\setlength\unitlength{1in}%
		\hspace*{\dimexpr0.5\paperwidth\relax}
		\makebox(0,-1)[c]{
			\begin{tabular}{c c}
				R.A. González and A.L. Cedeño,
				Simultaneous Input and State Estimation under Output Quantization: A Gaussian Mixture approach, \\
				accepted to IEEE Control Systems Letters, uploaded to ArXiv on July 4th, 2025 \\
\end{tabular}}}}

\title{\LARGE \bf
	Simultaneous Input and State Estimation under Output Quantization: A Gaussian Mixture approach
}

\author{Rodrigo A. González$^{1,*}$ and Angel L. Cedeño$^2$ 
	\thanks{ ${ }^1$Dept. of Mechanical Engineering, Eindhoven University of Technology, The Netherlands. ${ }^2$University of Santiago of Chile (USACH), Faculty of Engineering, Electrical Engineering Department, Chile. This work was supported in part by the grant ANID-Fondecyt 3240181, by the ANID-Basal Project AFB240002 (AC3E), and by E2TECH, University of Santiago of Chile. ${ }^*$Corresponding author, Email: \tt r.a.gonzalez@tue.nl.
	}%
}

\begin{document}
	
	\maketitle
	\thispagestyle{empty}
	\pagestyle{empty}

\begin{abstract}
	Simultaneous Input and State Estimation (SISE) enables the reconstruction of unknown inputs and internal states in dynamical systems, with applications in fault detection, robotics, and control. While various methods exist for linear systems, extensions to systems with output quantization are scarce, and no formal connections to limit Kalman filters are known in this context. This work addresses these gaps by proposing a novel SISE algorithm for linear systems with quantized output measurements. The proposed algorithm introduces a Gaussian mixture model formulation of the observation model, which leads to closed-form recursive equations in the form of a Gaussian sum filter. In the absence of input prior knowledge, the recursions are shown to converge to a limit-case SISE algorithm, implementable as a bank of linear SISE filters running in parallel. A simulation example is presented to illustrate the effectiveness of the proposed approach.
\end{abstract}
\begin{keywords}
	Simultaneous input and state estimation, Bayesian filtering, Quantized systems
\end{keywords}
\section{Introduction}
Simultaneous Input and State Estimation (SISE) refers to techniques that estimate both the internal states of dynamical systems and unknown inputs, such as disturbances or external forces \cite{Hieu2011}. Unlike classical state estimation methods that require full input knowledge, SISE is proposed for scenarios where some inputs are only partially known, or where nonlinear dynamics or unmodeled external effects are treated as unmeasured inputs that must be estimated. SISE algorithms have been applied in a wide range of areas, including fault detection \cite{Yao2022} and control of mechanical systems \cite{Valikhani2019}.

Extensive research has been conducted on SISE for linear time-invariant (LTI) systems. Early work focused on input estimation \cite{glover1969linear}, while \cite{kitanidis1987unbiased} proposed a recursive state estimator when the inputs are unknown. The work in \cite{gillijns2007unbiased} extended the aforementioned results to joint minimum-variance unbiased input and state estimation, leading to further work on statistical optimality \cite{hsieh2009extension, yin2025simultaneous}. The algorithm in \cite{gillijns2007unbiased} was later reinterpreted as a limiting form of the Kalman filter, obtained as the input covariance tends to infinity \cite{bitmead2019kalman}. This insight has inspired extensions for systems with direct feedthrough \cite{Song2024} and for smoothing algorithms \cite{gakis2024limit}.

Extending SISE to nonlinear systems poses greater challenges due to the limitations of standard Kalman-based techniques. While effective for state estimation, the extended Kalman filter is not readily applicable for SISE, since the system description remains unspecified when linearizing nonlinear functions with coupled states and inputs. A Bayesian approach has been proposed to address this issue \cite{Fang2013}, but the method requires the nonlinear functions to be continuously differentiable. This condition is not satisfied by many important nonlinearities, such as quantization, which is common in communication channels and encoder sensing.

Although state estimation with complete input knowledge under output quantization has received attention \cite{cedeno2021, Cedeno2021b}, SISE in this setting remains underexplored. Existing works typically focus on specific classes of nonlinear systems \cite{kim2020simultaneous, tian2024composite} and cannot directly handle output quantization. A notable exception is \cite{Zou2020}, which proposed a moving horizon estimator for dynamically quantized systems. While promising, this work neglects noise statistics and reduces to treating quantization error as noise when the quantization is static. Moreover, the optimization problem solved at each time step does not directly provide real-time estimates of the current system state, but rather of a delayed version of it. 

To date, no SISE algorithm accounts for the noise statistics in the presence of output quantization. This work addresses this gap in the literature through the following contributions:
\begin{enumerate}[label=C\arabic*]
	\item We propose a Bayesian filtering approach to simultaneously estimate the inputs and the states of a dynamical system subject to output quantization. To this end, 
	\begin{enumerate}[label=\alph*)]
		\item we describe the observation model as a Gaussian mixture, obtained by approximating the output likelihood function using a Gauss-Legendre quadrature rule; and
		\item we derive a Gaussian sum filter (GSF) for SISE by propagating the Bayesian filtering equations.       
	\end{enumerate}
	\item As the input covariance tends to infinity, we prove that the approach in Contribution C1 yields a SISE algorithm with zero incorporated prior input knowledge that can be interpreted as a bank of LTI SISE algorithms of the type introduced in \cite{gillijns2007unbiased} running in parallel.
	\item We illustrate the effectiveness of the proposed approach via a simulation example. 
\end{enumerate}

The remainder of this letter is organized as follows. In Section \ref{sec:problemformulation} we describe the system setup and the probabilistic framework considered for SISE. Section \ref{sec:gsf} contains the key contribution of this work, namely, the GSF for SISE considering quantized outputs.  Implementation aspects and the connection to the LTI SISE algorithm in \cite{gillijns2007unbiased} are discussed in Section~\ref{sec:implementation}. A simulation example is given in Section \ref{sec:simulations}, and Section \ref{sec:conclusions} concludes this letter.

\textit{Notation}: Vectors and matrices are represented in bold. Unless otherwise indicated, all vectors are assumed to be column vectors, and the identity matrix of appropriate dimensions is denoted by $\mathbf{I}$. For a collection of vectors $\mathbf{x}_i$, where $i = 1, \dots, K$, we use the notation $\mathbf{x}_{1:K}$ to denote the set $\{\mathbf{x}_1, \dots, \mathbf{x}_K\}$. The notation $\mathcal{N}(\mathbf{x}, \bm{\mu}, \bm{\Sigma})$ denotes a Gaussian distribution of the random variable $\mathbf{x}$ with mean $\bm{\mu}$ and covariance matrix $\bm{\Sigma}$. 

\section{Setup and problem formulation}
\label{sec:problemformulation}
\subsection{System setup}
\begin{figure}
	\centering
	\includegraphics[width=0.99\linewidth]{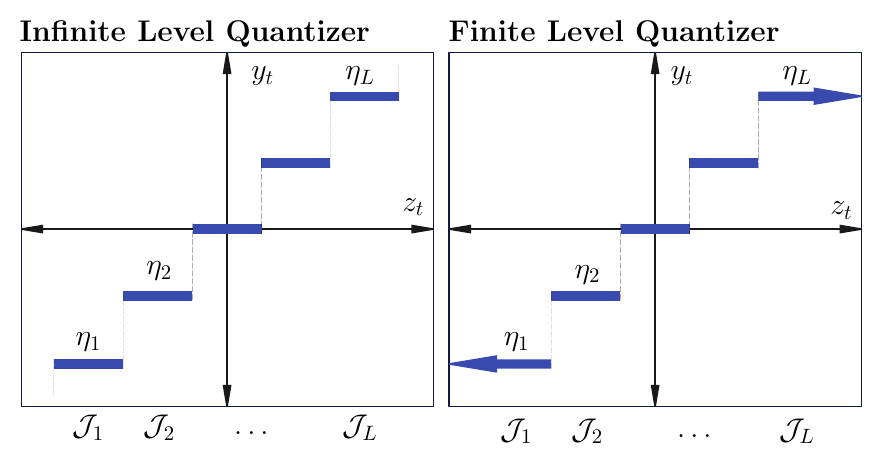}
	\vspace{-0.75cm}
	\caption{Univariate quantizer functions. Left: Quantizer with an infinite number of quantization levels. Right: Quantizer with $L$ quantization levels. In this case, the sets $\mathcal{J}_1$ and $\mathcal{J}_L$ are semi-infinite intervals. }
	\label{fig:quantizer_diagram}
	\vspace{-0.3cm}
\end{figure}

Consider the following discrete-time, multi-input, multi-output system with unknown input excitation:
\begin{align}
	\mathbf{x}_{t+1}&=\mathbf{Ax}_{t}+\mathbf{G}\mathbf{d}_{t}+\mathbf{w}_{t},\label{eqn:general_system}\\
	\mathbf{z}_{t}&=\mathbf{Cx}_{t}+\mathbf{v}_{t},\label{eqn:general_system2}\\
	\mathbf{y}_{t}&=\mathfrak{q}(\mathbf{z}_{t}),\label{eqn:general_system3}
\end{align}
where $\mathbf{d}_{t} \in \mathbb{R}^m$, $\mathbf{x}_{t} \in \mathbb{R}^{n}$, $\mathbf{z}_{t} \in \mathbb{R}^p$, and $\mathbf{y}_{t} \in \mathbb{R}^p$, are the unknown system input, the state vector, the linear output (not physically measurable), and the measured output, respectively. The system matrices have dimensions $\mathbf{A} \in \mathbb{R}^{n\times n}$, $\mathbf{G} \in \mathbb{R}^{n\times m}$, and $\mathbf{C} \in \mathbb{R}^{p\times n}$. No direct feedthrough term is included in this paper for simplicity, as the non-zero direct feedthrough case requires an additional input estimation step \cite{yong2016unified}. The process noise $\mathbf{w}_{t} \in \mathbb{R}^{n}$ and linear output noise $\mathbf{v}_{t} \in \mathbb{R}^p$ are jointly independent Gaussian-distributed stochastic processes, with zero mean and covariance matrices $\mathbf{Q}$ and $\mathbf{R}$, respectively. The initial condition $\mathbf{x}_1$ is also Gaussian distributed, of mean $\bm{\mu}_1$ and covariance $\mathbf{P}_1$. To ensure that all inputs $\mathbf{d}_{t-1}$ are observable from the output $\mathbf{y}_t$, we assume the following rank condition:
\begin{assumption}[Assump. 1, \cite{gillijns2007unbiased}]
	\label{assumption0}
	$\text{rank}(\mathbf{CG})=m$.
\end{assumption}
The measured output is a nonlinear function of $\mathbf{z}_t$, characterized by a piecewise constant function $\mathfrak{q}(\cdot)$ given by
\begin{equation}\label{eqn:quantizer}
	\mathfrak{q}(\mathbf{z}_{t})=\left\lbrace\begin{matrix}
		\bm{\eta}_1 & \textrm{ if } & \mathbf{z}_t \in \mathcal{J}_1,\\
		\bm{\eta}_2 & \textrm{ if } & \mathbf{z}_t \in \mathcal{J}_2,\\
		\vdots\\
		\bm{\eta}_L & \textrm{ if } & \mathbf{z}_t \in \mathcal{J}_L,\\
	\end{matrix}\right.
\end{equation}
where the sets $\mathcal{J}_i\subset \mathbb{R}^{p}$ are disjoint hyperrectangles that collectively partition $\mathbb{R}^p$. As $L \to \infty$, the function can represent an infinite-level multivariate quantizer, while for a finite $L$, it corresponds to a finite-level quantizer. Examples of these two scenarios are shown in Figure \ref{fig:quantizer_diagram}. In the extreme case of $L = 2$ and $p = 1$, $\mathfrak{q}(\mathbf{z}_{t})$ reduces to a binary quantizer.
\subsection{Probabilistic description}
\label{subsec:probabilistic}
To estimate the state and input given data, we adopt a Bayesian filtering approach \cite{sarkka2013bayesian}. After defining the extended state vector $\tilde{\mathbf{x}}_t := [\mathbf{x}_t^\top, \mathbf{d}_{t-1}^\top]^\top$, the Bayesian framework for SISE requires expressions for the joint state transition model $p(\tilde{\mathbf{x}}_{t+1}|\tilde{\mathbf{x}}_{t})$, as well as the PDF $p(\mathbf{y}_{t} | \tilde{\mathbf{x}}_{t})$ of the observation model. Since the SISE algorithm for linear systems (e.g., \cite{gillijns2007unbiased}) can be viewed as a limiting case of Kalman filtering \cite{bitmead2019kalman}, we exploit this connection to first model the unknown input $\mathbf{d}_t$. This model is later generalized in Section \ref{subsec:sisegeneral} by tending the input covariance to infinity.
\begin{assumption}
	\label{assumption1}
	The unknown input sequence is drawn from a set of independent and identically distributed (i.i.d.) Gaussian random variables with mean $\mathfrak{d}$ and covariance $\mathbf{D}$, and is independent of the noise sequence $\mathbf{w}_t$.
\end{assumption}

Under Assump. \ref{assumption1}, the joint state update is obtained as
\begin{equation}
	\label{stateupdate}
	p(\tilde{\mathbf{x}}_{t+1}|\tilde{\mathbf{x}}_{t}) =\mathcal{N}(\tilde{\mathbf{x}}_{t+1}; \tilde{\mathbf{A}}\tilde{\mathbf{x}}_t + \tilde{\mathbf{b}}, \tilde{\mathbf{Q}}),
\end{equation}
with the block matrices
\begin{equation}
	\tilde{\mathbf{A}} \hspace{-0.05cm} = \hspace{-0.08cm}\begin{bmatrix}
		\mathbf{A}\hspace{-0.05cm} & \hspace{-0.05cm}\mathbf{0} \\
		\mathbf{0}\hspace{-0.05cm} & \hspace{-0.05cm}\mathbf{0}
	\end{bmatrix}\hspace{-0.03cm}, \hspace{0.1cm} \tilde{\mathbf{b}} \hspace{-0.05cm}=\hspace{-0.08cm} \begin{bmatrix}
		\mathbf{G}\mathfrak{d} \\
		\mathfrak{d}
	\end{bmatrix} \hspace{-0.03cm}, \hspace{0.1cm} \tilde{\mathbf{Q}} \hspace{-0.05cm}=\hspace{-0.08cm} \begin{bmatrix}
		\mathbf{Q} \hspace{-0.05cm} + \hspace{-0.05cm} \mathbf{GDG}^\top & \hspace{-0.05cm}\mathbf{GD} \\
		\mathbf{DG}^\top & \hspace{-0.05cm}\mathbf{D}
	\end{bmatrix}. \notag 
\end{equation}
To obtain $p(\mathbf{y}_t|\tilde{\mathbf{x}}_t)$, it is possible to relate the discrete-type random variable $\mathbf{y}_t$ to the continuous-type random variable $\mathbf{z}_t$ by noting that the probability that $\mathbf{y}_t$ takes a particular value $\bm{\eta}_i$ is equal to the probability that $\mathbf{z}_t\in\mathcal{J}_i$, i.e.,
\begin{equation}
	\label{pmf}
	p(\mathbf{y}_{t}|\tilde{\mathbf{x}}_{t}) = \int_{\mathcal{J}_{i}} \mathcal{N}(\mathbf{z}_t;\tilde{\mathbf{C}}\tilde{\mathbf{x}}_{t},\mathbf{R})\textnormal{d}\mathbf{z}_t,    
\end{equation}
where $\tilde{\mathbf{C}}=[\mathbf{C}, \hspace{0.06cm}\mathbf{0}]$. These integrals do not have explicit expressions in general, which means that the observation model does not have a closed form. The key idea is to approximate such integrals, resulting in a finite sum expression that explicitly depends on $\tilde{\mathbf{x}}_{t}$. There are various numerical methods available for this purpose, such as Monte Carlo techniques, Newton-Cotes formulas, and Gaussian quadrature. In this work, we employ the Gauss-Legendre quadrature rule \cite{davis2007methods}, which evaluates the integrand at specific quadrature points, weighting each accordingly, yielding an approximation of $p(\mathbf{y}_{t}|\tilde{\mathbf{x}}_{t})$ as a weighted sum of quadratic exponential functions in $\tilde{\mathbf{x}}_{t}$. More precisely,
\begin{equation}
	\label{gmmy}
	p(\mathbf{y}_t|\tilde{\mathbf{x}}_t) \approx \sum_{\kappa=1}^{K} \varphi_t^{\kappa}\mathcal{N}(\zeta_t^{\kappa}(\mathbf{y}_t);\tilde{\mathbf{C}}\tilde{\mathbf{x}}_t + \bm{\mu}_t^\kappa,\mathbf{R}),
\end{equation}
where $\varphi_t^{\kappa}$, $\zeta_t^{\kappa}(\cdot)$ and $\bm{\mu}_t^\kappa$ depend on the nonlinearity $\mathfrak{q}$, i.e., on the quantizer levels and sets. Closed-form expressions of these quantities can be found in Table 1 of \cite{Cedeno2021b} for the scalar output case, where, in the notation of \cite{Cedeno2021b}, $\varphi_t^{\kappa}$ corresponds to $\varsigma_t^{\tau}$, $\zeta_t^{\kappa}$ to $\eta_t^{\tau}$, and $\bm{\mu}_t^{\kappa}$ to $\mu_t^{\tau}$.


\subsection{Problem formulation}
\vspace{-0.1cm}
Given the system \eqref{eqn:general_system}-\eqref{eqn:general_system3} and the probabilistic model in Section \ref{subsec:probabilistic}, our goal is to derive a method for computing the state filtering PDF $p(\mathbf{x}_t|\mathbf{y}_{1:t})$ and the input filtering PDF $p(\mathbf{d}_{t-1}|\mathbf{y}_{1:t})$ using only the quantized output data $\mathbf{y}_1,\mathbf{y}_2,\dots,\mathbf{y}_t$, with $t\in \mathbb{N}$. Once these PDFs are obtained via a recursive algorithm, we analyze their behavior when $\mathbf{D}^{-1}\to \mathbf{0}$, i.e., the case where a non-informative prior is imposed on the system input.

\section{Gaussian sum filtering for simultaneous input and state estimation under quantization}
\label{sec:gsf}
A core property of the probabilistic description of the system using \eqref{stateupdate} and \eqref{gmmy} is that the resulting PDFs allow for the direct derivation of a GSF for simultaneous input and state estimation. Subsection \ref{subsec:sisegaussian} introduces this filtering procedure, while such filter is later extended to inputs with uninformative prior knowledge in Subsection \ref{subsec:sisegeneral}, constituting the main contributions of this letter.

\subsection{SISE under Gaussian assumption on unknown input}
\label{subsec:sisegaussian}
Theorem \ref{thm:filtering} presents the filtering procedure for SISE with quantized outputs under the Gaussianity and independence assumption on the unknown input sequence (Assumption \ref{assumption1}).

\begin{theorem}\label{thm:filtering}
	Consider the system in \eqref{eqn:general_system}-\eqref{eqn:general_system3}, and the PDF $p(\mathbf{y}_t|\mathbf{x}_t)$ in \eqref{gmmy}, assuming Assumptions \ref{assumption0} and \ref{assumption1} hold. Then, the GSF for state-space systems with quantized outputs and unknown inputs is given as follows. For time $t=1$, the PDFs of the time update step are given by $p(\mathbf{x}_1)=\mathcal{N}(\mathbf{x}_1;\bm{\mu}_1,\mathbf{P}_1)$ and $p(\mathbf{d}_0)=\mathcal{N}(\mathbf{d}_0;\mathfrak{d},\mathbf{D})$. For $t=2,3,\dots$, the following steps are defined:\\
	
	\noindent \textbf{Measurement Update:} The filtered PDFs of the current state, $\mathbf{x}_t$, and the previous sample of the unknown input, $\mathbf{d}_{t-1}$, given measurements of the quantized output $\mathbf{y}_1,\dots,\mathbf{y}_t$, are the Gaussian mixture models (GMMs)
	\begin{align}\label{eqn:correction_filtering} 
		p(\mathbf{x}_t|\mathbf{y}_{1:t})&=\sum_{k=1}^{M_{t|t}} \delta_{t|t}^{k}\mathcal{N}(\mathbf{x}_{t};\hat{\mathbf{x}}_{t|t}^{k},\bm{\Sigma}_{t|t}^{k}), \\
		p(\mathbf{d}_{t-1}|\mathbf{y}_{1:t})&=\sum_{k=1}^{M_{t|t}} \delta_{t|t}^{k}\mathcal{N}(\mathbf{d}_{t-1};\hat{\mathbf{d}}_{t-1|t}^{k},\bm{\Gamma}_{t-1|t}^{k}),
	\end{align}
	where $M_{t|t}=KM_{t|t-1}$, and for each pair $(\kappa,\ell)$, with $\kappa=1,\dots,K$ and $\ell=1,\dots, M_{t|t-1}$, a new index $k=(\ell-1)K+\kappa$ is defined such that the GMMs are described by
	\begin{align}
		\delta_{t|t}^{k} & \!=\! \bar{\delta}_{t|t}^{k}/\textstyle\sum_{r=1}^{M_{t|t}}\bar{\delta}_{t|t}^{r}, \label{eqn:lemma_filtering_1}\\	
		\bar{\delta}_{t|t}^{k}&\!=\!\varphi_t^{\kappa} \delta_{t|t\hspace{-0.02cm}-\hspace{-0.02cm}1}^{\ell}\mathcal{N}\hspace{-0.08cm}\prt{\hspace{-0.03cm}\zeta_t^{\kappa}\hspace{-0.02cm}(\mathbf{y}_t);\hspace{-0.02cm}\mathbf{C}\hat{\mathbf{x}}_{t|t\hspace{-0.02cm}-\hspace{-0.02cm}1}^{\ell}\hspace{-0.09cm}+\hspace{-0.06cm}\bm{\mu}_t^\kappa\hspace{-0.02cm},\mathbf{R}\hspace{-0.06cm}+\hspace{-0.06cm}\mathbf{C}\bm{\Sigma}_{t|t\hspace{-0.01cm}-\hspace{-0.01cm}1}^{\ell}\hspace{-0.02cm}\mathbf{C}^{\hspace{-0.03cm}\top}\hspace{-0.04cm}}\hspace{-0.06cm},\label{eqn:lemma_filtering_2}\\
		\mathbf{K}_t^{\ell}&\!=\!\bm{\Sigma}_{t|t-1}^{\ell}\mathbf{C}^{\top}(\mathbf{R}\hspace{-0.05cm}+\hspace{-0.05cm}\mathbf{C}\bm{\Sigma}_{t|t-1}^{\ell}\mathbf{C}^{\top})^{-1}, \label{eqn:lemma_filtering_3}\\ 
		\hat{\mathbf{x}}_{t|t}^{k}&\!=\!\hat{\mathbf{x}}_{t|t-1}^{\ell}+\mathbf{K}_t^{\ell}(\zeta_t^{\kappa}(\mathbf{y}_t)-\mathbf{C}\hat{\mathbf{x}}_{t|t-1}^{\ell}-\bm{\mu}_t^\kappa),\label{eqn:lemma_filtering_4}\\
		\bm{\Sigma}_{t|t}^{k}&\!=\!(\mathbf{I}-\mathbf{K}_t^{\ell}\mathbf{C})\bm{\Sigma}_{t|t-1}^{\ell},\label{eqn:lemma_filtering_5} \\
		\mathbf{L}_t^{\ell}&\!=\!\mathbf{DG}^\top\mathbf{C}^{\top}(\mathbf{R}\hspace{-0.05cm}+\hspace{-0.05cm}\mathbf{C}\bm{\Sigma}_{t|t-1}^{\ell}\mathbf{C}^{\top})^{-1}, \label{eqn:lemma_filtering_6}  \\
		\hat{\mathbf{d}}_{t-1|t}^{k} &\!=\! \mathfrak{d} + \mathbf{L}_t^{\ell} (\zeta_t^{\kappa}(\mathbf{y}_t)-\mathbf{C}\hat{\mathbf{x}}_{t|t-1}^{\ell}-\bm{\mu}_t^\kappa), \label{eqn:lemma_filtering_8} \\
		\bm{\Gamma}_{t-1|t}^{k}&\!=\!(\mathbf{I}-\mathbf{L}_t^{\ell}\mathbf{CG})\mathbf{D},\label{eqn:lemma_filtering_9}            
	\end{align}
	where $K$, $\varphi_t^{\kappa}$, $\zeta_t^{\kappa}(\cdot)$ and $\bm{\mu}_t^\kappa$ are obtained from the GMM in \eqref{gmmy}. The initial values of the recursion are $M_{1|0}=1$, $\delta_{1|0}=1$, $\hat{\mathbf{x}}_{1|0}=\bm{\mu}_1$, and $\bm{\Sigma}_{1|0}=\mathbf{P}_1$.\\
	
	\noindent \textbf{Time Update:} The predicted PDF of the state $\mathbf{x}_{t+1}$ given measurements of the nonlinear output $\mathbf{y}_1,\dots,\mathbf{y}_t$ is given by
	\begin{equation}\label{eqn:prediccion_filtering}
		p(\mathbf{x}_{t+1}|\mathbf{y}_{1:t})=\sum_{k=1}^{M_{t+1|t}} \delta_{t+1|t}^{k}\mathcal{N}(\mathbf{x}_{t+1};\hat{\mathbf{x}}_{t+1|t}^{k},\bm{\Sigma}_{t+1|t}^{k}),
	\end{equation}
	where $M_{t+1|t}=M_{t|t}, \delta_{t+1|t}^{k}=\delta_{t|t}^{k}$, and
	\begin{align}
		\hat{\mathbf{x}}_{t+1|t}^{k}&=\mathbf{A}\hat{\mathbf{x}}_{t|t}^{k}+\mathbf{G}\mathfrak{d}, \label{eqn:lemma_filtering_10}\\	
		\bm{\Sigma}_{t+1|t}^{k}&=\mathbf{Q}+\mathbf{GDG}^\top+\mathbf{A}\bm{\Sigma}_{t|t}^{k}\mathbf{A}^\top. \label{eqn:lemma_filtering_11}
	\end{align}
\end{theorem}

\textit{Proof}: See Appendix A. \hfill $\square$

\vspace{-0.1cm}
\subsection{SISE under no input prior knowledge}
\label{subsec:sisegeneral}
Theorem \ref{thm:filtering} presents a recursive algorithm for SISE under quantized measurements, assuming specific input signal properties. Following \cite{bitmead2019kalman} for SISE with linear measurements, a SISE algorithm with no prior assumption on the input distribution emerges from Theorem \ref{thm:filtering} as the covariance $\mathbf{D}$ tends to infinity. The resulting filter, formalized in Theorem \ref{thm:filtering2}, is a GSF for SISE with quantized output measurements and a non-informative prior on the unknown input signal.

\begin{theorem}\label{thm:filtering2}
	Consider the system in \eqref{eqn:general_system}-\eqref{eqn:general_system3}, and the PDF $p(\mathbf{y}_t|\mathbf{x}_t)$ in \eqref{gmmy}, assuming Assumption \ref{assumption0} holds. Then, the GSF for state-space systems with quantized outputs and unknown inputs under an uninformative prior is given as follows. For time $t=1$, $p(\mathbf{x}_1)=\mathcal{N}(\mathbf{x}_1;\bm{\mu}_1,\mathbf{P}_1)$, and for $t=2,3,\dots$, the filtered PDFs of the current state, $\mathbf{x}_t$, and the previous sample of the unknown input, $\mathbf{d}_{t-1}$, given measurements of the quantized output $\mathbf{y}_{1:t}$ are the GMMs
	\begin{align}
		\label{pdfx}
		p(\mathbf{x}_t|\mathbf{y}_{1:t})&=\sum_{k=1}^{M_{t}} \delta_{t}^{k}\mathcal{N}(\mathbf{x}_{t};\hat{\mathbf{x}}_{t}^{k},\bm{\Sigma}_{t}^{k}), \\
		\label{pdfd}
		p(\mathbf{d}_{t-1}|\mathbf{y}_{1:t})&=\sum_{k=1}^{M_{t}} \delta_{t}^{k}\mathcal{N}(\mathbf{d}_{t-1};\hat{\mathbf{d}}_{t-1}^{k},\bm{\Gamma}_{t-1}^{k}),
	\end{align}
	where $M_{t}=KM_{t-1}$, and for each pair $(\kappa,\ell)$, with $\kappa=1,\dots,K$ and $\ell=1,\dots, M_{t-1}$, a new index $k=(\ell-1)K+\kappa$ is defined such that the GMMs are described by
	\begin{align}
		\delta_{t}^{k} & \!=\! \bar{\delta}_{t}^{k}/\textstyle\sum_{r=1}^{M_{t}}\bar{\delta}_{t}^{r}, \label{delta1} \\	
		\bar{\delta}_{t}^{k} & \!=\! \varphi_t^{\kappa} \delta_{t-1}^{\ell}\sqrt{\frac{\det(\bm{\Gamma}_{t-1}^{k})}{\det(\mathbf{R}\hspace{-0.05cm}+\hspace{-0.05cm}\mathbf{C}\mathbf{X}^{\ell}_{t}\mathbf{C}^{\top})}}, \label{delta2} \\	
		\mathbf{X}^{\ell}_{t}&=\mathbf{Q} + \mathbf{A}\bm{\Sigma}_{t-1}^\ell \mathbf{A}^\top, \label{xellt} \\
		\mathbfcal{L}_t^{\ell}&\!=\![\mathbf{G}^\top \mathbf{C}^\top(\mathbf{R}\hspace{-0.05cm}+\hspace{-0.05cm}\mathbf{C}\mathbf{X}^{\ell}_{t}\mathbf{C}^{\top})^{-1}\mathbf{CG} ]^{-1} \notag \\
		&\times \mathbf{G}^\top \mathbf{C}^\top (\mathbf{R}\hspace{-0.05cm}+\hspace{-0.05cm}\mathbf{C}\mathbf{X}^{\ell}_{t}\mathbf{C}^{\top})^{-1},  \label{ltell} \\
		\mathbfcal{J}_t^\ell&=\mathbf{X}^{\ell}_{t}\mathbf{C}^\top (\mathbf{R}+\mathbf{C}\mathbf{X}^{\ell}_{t}\mathbf{C}^\top)^{-1}, \label{jtell} \\
		\hat{\mathbf{d}}_{t\hspace{-0.02cm}-\hspace{-0.02cm}1}^{k} \hspace{-0.03cm}&\!=\! \mathbfcal{L}_t^{\ell} (\zeta_t^{\kappa}(\mathbf{y}_t)-\mathbf{C}\mathbf{A}\hat{\mathbf{x}}_{t-1}^{\ell}-\bm{\mu}_t^\kappa), \label{dt1} \\
		\hat{\mathbf{x}}_{t}^{k}&\!=\!\hspace{-0.04cm}\mathbf{A}\hat{\mathbf{x}}_{t\hspace{-0.025cm}-\hspace{-0.025cm}1}^{\ell}\hspace{-0.11cm}+\hspace{-0.07cm}\mathbf{G}\hat{\mathbf{d}}_{t\hspace{-0.02cm}-\hspace{-0.02cm}1}^{k}\hspace{-0.12cm}+\hspace{-0.08cm}\mathbfcal{J}_t^\ell \hspace{-0.03cm}(\hspace{-0.01cm}\zeta_t^{\kappa}\hspace{-0.02cm}(\mathbf{y}_{\hspace{-0.02cm}t}\hspace{-0.01cm})\hspace{-0.09cm}-\hspace{-0.1cm}\bm{\mu}_t^{\hspace{-0.02cm}\kappa}\hspace{-0.11cm}-\hspace{-0.09cm}\mathbf{C}\mathbf{A}\hat{\mathbf{x}}_{t\hspace{-0.025cm}-\hspace{-0.025cm}1}^{\ell}\hspace{-0.12cm}-\hspace{-0.09cm}\mathbf{CG}\hat{\mathbf{d}}_{t\hspace{-0.025cm}-\hspace{-0.025cm}1}^{k}\hspace{-0.02cm})\hspace{-0.01cm}, \label{xtsise} \\
		\bm{\Gamma}_{t\hspace{-0.02cm}-\hspace{-0.02cm}1}^{k}&\hspace{-0.03cm}\!=\![\mathbf{G}^\top \mathbf{C}^\top(\mathbf{R}+\mathbf{C}\mathbf{X}^{\ell}_{t}\mathbf{C}^{\top})^{-1}\mathbf{CG}]^{-1}, \\
		\bm{\Sigma}_{t}^{k}&\!=\!(\mathbf{I}-\mathbfcal{J}_t^{\ell}\mathbf{C})(\mathbf{X}^{\ell}_{t}- \mathbf{G}\mathbfcal{L}^{\ell}_{t}\mathbf{C}\mathbf{X}^{\ell}_{t}+\mathbf{G}\bm{\Gamma}_{t-1}^{k}\mathbf{G}^{\top}), \label{sigmat}     
	\end{align}
	where $K$, $\varphi_t^{\kappa}$, $\zeta_t^{\kappa}(\cdot)$ and $\bm{\mu}_t^\kappa$ are obtained from the GMM in \eqref{gmmy}. The initial values of the recursion are $M_{1|0}=1$, $\delta_{1|0}=1$, $\hat{\mathbf{x}}_{1|0}=\bm{\mu}_1$, and $\bm{\Sigma}_{1|0}=\mathbf{P}_1$.
\end{theorem}

\textit{Proof}: See Appendix B. \hfill $\square$

\section{Implementation and discussion}
\label{sec:implementation} 
\subsection{Implementation: Gaussian sum reduction}
The number of Gaussian components during the filtering process grows exponentially according to the update law $M_{t|t}=KM_{t|t-1}$ in Theorem \ref{thm:filtering} and $M_{t}=KM_{t-1}$ in Theorem \ref{thm:filtering2}, for both input and state estimation. Thus, more compact representations are necessary to maintain tractable computational loads. Several approaches exist for Gaussian sum reduction for minimizing information loss \cite{kitagawa1994two}. For instance, probabilistic merging combines components based on statistical similarity, and the integrated squared error optimization method minimizes the distance between the original and reduced mixtures through a variational formulation. In this work, we adopt the mixture reduction framework in \cite{runnalls2007kullback}, which optimally reduces the mixture by minimizing the Kullback-Leibler divergence between the full and reduced models, offering an effective balance between computational efficiency and approximation accuracy. This reduction is performed jointly for the state and input GMMs, ensuring that the resulting weights remain consistent across both filtering distributions for the subsequent recursion.

\subsection{Parallel implementation and relation to LTI SISE methods}

\begin{figure*}
	\centering
	\includegraphics[width=1\linewidth]{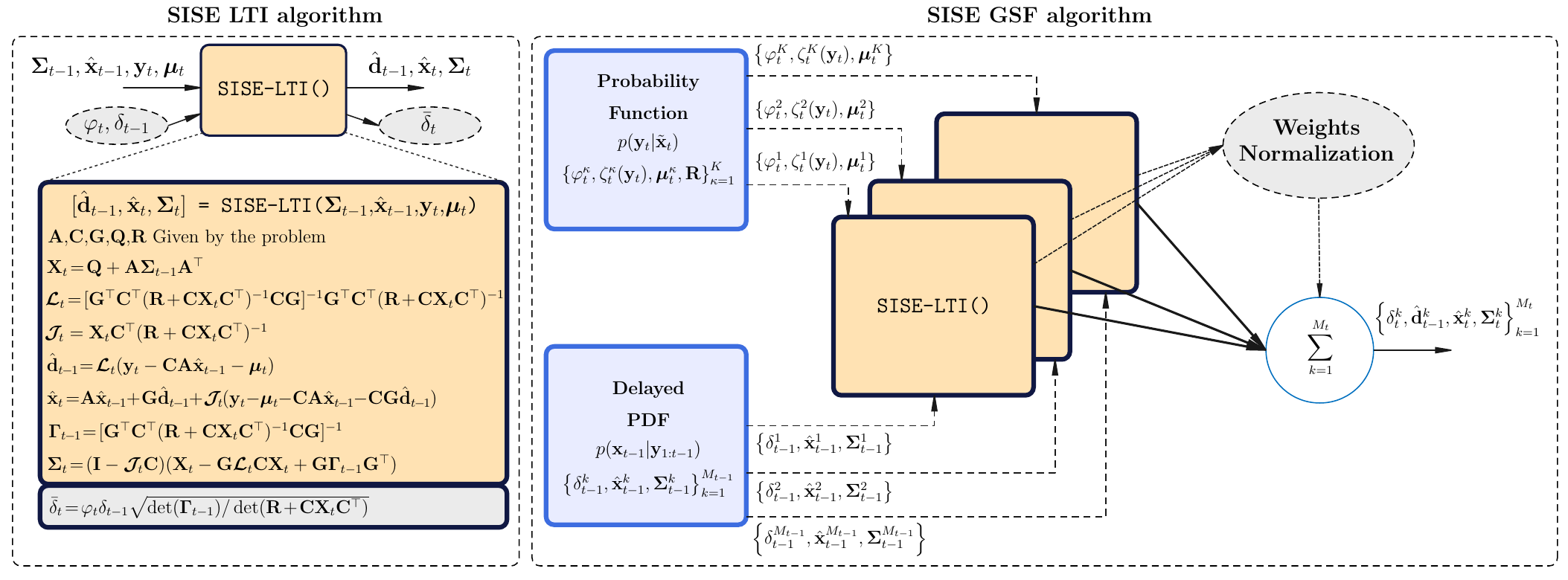}
	\vspace{-0.7cm}
	\caption{One iteration of the SISE LTI algorithm in \cite{gillijns2007unbiased} (left) and one recursion of the algorithm in Thm. 2 (right). Each iteration of the SISE-GSF can be interpreted as a bank of LTI SISE iterations running in parallel, followed by a weight normalization to compute the state and input filtering GMMs.}
	\label{fig:parallel_sise_diagram}
	\vspace{-0.4cm}
\end{figure*}

A important aspect of the proposed algorithms in Theorems \ref{thm:filtering} and \ref{thm:filtering2} is the inherent parallelism of the recursive equations. This implementation feature is illustrated in Figure~\ref{fig:parallel_sise_diagram}. Specifically, the algorithm performs a separate LTI SISE update for each Gaussian term, where each update is fully determined by the GMM representation from the previous time instant. All these iterations can be performed concurrently on dedicated hardware, enabling efficient computation. Each iteration step, portrayed on the left-hand side of Figure \ref{fig:parallel_sise_diagram}, corresponds exactly with the LTI SISE method found in e.g., \cite{gillijns2007unbiased} and analyzed in \cite{abooshahab2022simultaneous}, where the covariance matrix expression $\bm{\Sigma}_t$ in \eqref{sigmat} corresponds to Eq. (7) of \cite{abooshahab2022simultaneous} after some algebraic manipulations. After each LTI SISE update, the resulting Gaussian component is weighted according to \eqref{delta2} and normalized, yielding the state and input GMMs.

\subsection{Extensions to other output nonlinearities}
The quantization nonlinearity, as well as any other static output nonlinearity, only acts upon the observation model $p(\mathbf{y}_t|\tilde{\mathbf{x}}_t)$, while the joint state update equation \eqref{stateupdate} remains invariant to such nonlinearities. Therefore, as long as the observation model can be approximated in the Gaussian mixture form of \eqref{gmmy}, for appropriately chosen functions $\varphi_t^{\kappa}$, $\zeta_t^{\kappa}(\cdot)$, and means $\bm{\mu}_t^\kappa$, the Gaussian sum filtering procedure outlined in Theorems \ref{thm:filtering} and \ref{thm:filtering2} can be directly applied to derive SISE algorithms. For instance, a GMM structure of the form \eqref{gmmy} can also be reached via a Gauss-Legendre approximation when the nonlinearity is given by a deadzone, saturation, or a linear rectifier \cite{cedeno2024gaussian}.

\section{Simulation example}\label{sec:simulations} 
This section compares the filters proposed in Theorems \ref{thm:filtering} and \ref{thm:filtering2}, denoted $\textnormal{SISE-GSF}_{1}$ and $\textnormal{SISE-GSF}_{2}$, respectively, with the method from \cite{gillijns2007unbiased}, referred to as SISE-LTI. In $\textnormal{SISE-GSF}_{2}$, the input and state estimates are computed as the means of the random variables characterized by the densities in \eqref{pdfd} and \eqref{pdfx}, respectively. For $\textnormal{SISE-GSF}_{1}$ we consider $\mathfrak{d}=0$ and $\mathbf{D}=1000$. The purpose of this simulation is to show the impact of quantization by comparing scenarios that explicitly account for it ($\textnormal{SISE-GSF}_{1,2}$) with one that does not (SISE-LTI). Consider the following state-space system
\begin{align}
	\mathbf{x}_{t+1}&=
	\begin{bmatrix}
		0.9&0\\
		0&0.7
	\end{bmatrix}\mathbf{x}_{t}+
	\begin{bmatrix}
		2\\1
	\end{bmatrix}u_{t}+\mathbf{w}_{t}, \notag \\
	z_{t}&=
	\begin{bmatrix}
		1.0 & 1.5
	\end{bmatrix} \mathbf{x}_{t}+v_{t}, \notag 
\end{align}
with the quantization function described by $y_{t} = \Delta \textnormal{round}(z_t/\Delta)$, where $\Delta=1,5,10,15$. The process and measurement noise terms are modeled as Gaussian distributions $\mathbf{w}_t \sim \mathcal{N}(\mathbf{w}_t;\mathbf{0}, 0.1\mathbf{I})$ and $v_t \sim \mathcal{N}(v_t;0, 0.1)$. The control input $u_t$ is sampled from $\mathcal{N}(u_t;0, 20)$, and the initial state is assumed to follow $\mathbf{x}_1 \sim \mathcal{N}(\mathbf{x}_1; [2,1]^{\top}, 0.5\mathbf{I})$. To approximate $p(\mathbf{y}_t|\tilde{\mathbf{x}}_t)$ in \eqref{gmmy}, we consider $K=20$. Note that, as discussed in \cite{Cedeno2021b} for Gaussian sum filtering with quantized output data, increasing $K$ results in an increment in computational time; however, this does not become prohibitive for moderate-dimensional systems when $10 \leq K \leq 100$.

\begin{figure}
	\centering
	\includegraphics[width=0.97\linewidth]{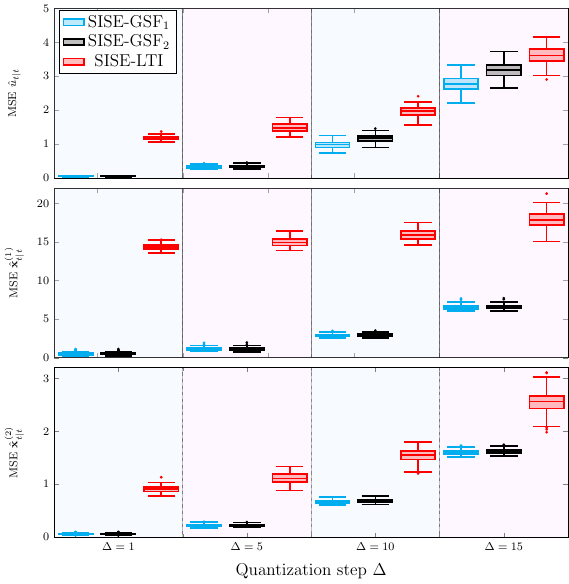}
	\vspace{-0.4cm}
	\caption{Box plots of the MSE between the true and estimated sequences obtained using the SISE-GSF and SISE-LTI algorithms, for different quantization step sizes. Top: input estimation. Middle and bottom: state estimation.}
	\label{fig:input_boxplot_mse}\vspace{-0.6cm}
\end{figure}

Figure~\ref{fig:input_boxplot_mse} shows box plots of the Mean Squared Error (MSE) between the true and estimated input and states sequences using the SISE-GSF and SISE-LTI algorithms, based on a Monte Carlo simulation with 100 realizations per quantization step size. It can be observed that the proposed SISE-GSF filters provide better input and state estimates, as they account for quantization in the data, unlike the SISE-LTI filter, which does not consider this effect. Additionally, it is evident that as the quantization step size increases, the estimation error also grows. This behavior is expected, since a more coarse quantization leads to greater information loss in the output, which in turn increases the estimation error.

\section{Conclusions}
\label{sec:conclusions}
In this paper, we have derived a recursive algorithm for SISE in linear systems under output quantization. The method is grounded in Bayesian principles, with the observation model represented as a GMM. Such description naturally delivers a Gaussian sum filter, from which a SISE algorithm with flat prior input information was derived. Such filtering procedure has been interpreted as several LTI SISE algorithms acting in parallel, enabling efficient and accurate computation. Future work includes extending the methodology to systems with direct feedthrough, deriving the smoothing algorithm for this setup, and analyzing its stability, convergence, and statistical properties.

\section*{Appendix}
\subsection{Proof of Theorem \ref{thm:filtering}}
\label{appendix:thm1}
From Theorem 2 of \cite{Cedeno2021b} for the known input case, the measurement update for the extended state when conditioned on the quantized output measurements is given by
\begin{equation}
	\label{totalpdf}
	p(\tilde{\mathbf{x}}_t|\mathbf{y}_{1:t}) = \sum_{k=1}^{M_{t|t}} \delta_{t|t}^{k}\mathcal{N}(\tilde{\mathbf{x}}_{t};\hat{\tilde{\mathbf{x}}}_{t|t}^{k},\tilde{\bm{\Sigma}}_{t|t}^{k}),
\end{equation}
where $M_{t|t}=KM_{t|t-1}$ , $\delta_{t|t}^{k}$ can be written only in terms of the system state as in \eqref{eqn:lemma_filtering_1} and \eqref{eqn:lemma_filtering_2}, and
\begin{align}
	\tilde{\mathbf{K}}_t^{\ell}&\!=\! \tilde{\bm{\Sigma}}_{t|t-1}^{\ell}\tilde{\mathbf{C}}^{\top}(\mathbf{R}\hspace{-0.05cm}+\hspace{-0.05cm}\tilde{\mathbf{C}}\tilde{\bm{\Sigma}}_{t|t-1}^{\ell}\tilde{\mathbf{C}}^{\top})^{-1}, \notag \\ 
	\hat{\tilde{\mathbf{x}}}_{t|t}^{k}&\!=\!\hat{\tilde{\mathbf{x}}}_{t|t-1}^{\ell}+\tilde{\mathbf{K}}_t^{\ell}(\zeta_t^{\kappa}(\mathbf{y}_t)-\mathbf{C}\hat{\tilde{\mathbf{x}}}_{t|t-1}^{\ell}-\bm{\mu}_t^\kappa), \notag \\
	\tilde{\bm{\Sigma}}_{t|t}^{k}&\!=\!(\mathbf{I}-\tilde{\mathbf{K}}_t^{\ell}\tilde{\mathbf{C}})\tilde{\bm{\Sigma}}_{t|t-1}^{\ell}.  \notag
\end{align}
After integrating \eqref{totalpdf} with respect to $\mathbf{d}_{t-1}$ and $\mathbf{x}_t$, the resulting marginals $p(\mathbf{x}_t|\mathbf{y}_{1:t})$ and $p(\mathbf{d}_{t-1}|\mathbf{y}_{1:t})$ are also GMMs. Each Gaussian component has a mean given by the upper or lower block vector of  $\hat{\tilde{\mathbf{x}}}_{t|t}^{k}$, and a covariance given by the upper or lower block diagonal matrix of $\tilde{\bm{\Sigma}}_{t|t}^{k}$, respectively \cite[Sec. 8.1.2]{petersen2012}. When expanding in terms of $\mathbf{x}_t$ and $\mathbf{d}_{t-1}$, the measurement update yields $\bar{\delta}_{t|t}^{k}$ as in \eqref{eqn:lemma_filtering_2}, and
\begin{align}
	\tilde{\mathbf{K}}_t^{\ell}&\!:=\! \begin{bmatrix}
		\mathbf{K}_t^{\ell} \\
		\mathbf{L}_t^{\ell}
	\end{bmatrix} \!=\! \begin{bmatrix}
		\bm{\Sigma}_{t|t-1}^{\ell} \\
		\mathbf{DG}^\top
	\end{bmatrix} \mathbf{C}^{\top}\hspace{-0.04cm}(\mathbf{R}\hspace{-0.04cm}+\hspace{-0.04cm}\mathbf{C}\bm{\Sigma}_{t|t-1}^{\ell}\mathbf{C}^{\top})^{-1}, \notag \\
	\bm{\Sigma}_{t|t}^{k}&\!=\! \bm{\Sigma}_{t|t-1}^{\ell}\!-\!\bm{\Sigma}_{t|t-1}^{\ell}\!\mathbf{C}^{\top}\!(\mathbf{R}\!+\!\mathbf{C}\bm{\Sigma}_{t|t-1}^{\ell}\mathbf{C}^{\top})^{\hspace{-0.03cm}-\hspace{-0.03cm}1} \hspace{-0.01cm}\mathbf{C}\bm{\Sigma}_{t|t-1}^{\ell}, \notag \\
	\bm{\Gamma}_{t-1|t}^{k}&\!=\! \mathbf{D}-\mathbf{D}\mathbf{G}^\top\mathbf{C}^{\top}(\mathbf{R}+\mathbf{C}\bm{\Sigma}_{t|t-1}^{\ell}\mathbf{C}^{\top})^{-1} \mathbf{CGD}, \notag
\end{align}
which can be rewritten as \eqref{eqn:lemma_filtering_3}, \eqref{eqn:lemma_filtering_6}, \eqref{eqn:lemma_filtering_5}, and \eqref{eqn:lemma_filtering_9}, respectively. On the other hand, the time update is given by
\begin{equation}
	p(\tilde{\mathbf{x}}_{t+1}|\mathbf{y}_{1:t})=\sum_{k=1}^{M_{t+1|t}} \delta_{t+1|t}^{k}\mathcal{N}(\tilde{\mathbf{x}}_{t+1};\hat{\tilde{\mathbf{x}}}_{t+1|t}^{k},\tilde{\bm{\Sigma}}_{t+1|t}^{k}), \notag 
\end{equation}
where $M_{t+1|t}=M_{t|t}, \delta_{t+1|t}^{k}=\delta_{t|t}^{k}$, $\hat{\tilde{\mathbf{x}}}_{t+1|t}^{k}=\tilde{\mathbf{A}}\hat{\tilde{\mathbf{x}}}_{t|t}^{k}+\tilde{\mathbf{b}}$, and $\tilde{\bm{\Sigma}}_{t+1|t}^{k}=\tilde{\mathbf{Q}}+\tilde{\mathbf{A}}\tilde{\bm{\Sigma}}_{t|t}^{k}\tilde{\mathbf{A}}^\top.$ These expressions directly yield \eqref{eqn:lemma_filtering_9} and \eqref{eqn:lemma_filtering_10} when considering the first component of $\tilde{\mathbf{x}}_t$, concluding the proof. \hfill $\square$

\vspace{-0.1cm}
\subsection{Proof of Theorem \ref{thm:filtering2}}
\label{appendix:thm2}
\vspace{-0.1cm}
The algorithm is derived by letting $\mathbf{D}^{-1}\hspace{-0.07cm}\to\hspace{-0.07cm} \mathbf{0}$ in Theorem \ref{thm:filtering}. Let $\bm{\Sigma}_{t\hspace{-0.02cm}-\hspace{-0.02cm}1}^\ell\hspace{-0.03cm} (\ell\hspace{-0.05cm}=\hspace{-0.05cm}1\hspace{-0.02cm},\dots,\hspace{-0.02cm} M_{t\hspace{-0.02cm}-\hspace{-0.02cm}1}$) be the covariance of each Gaussian component of $p(\mathbf{x}_{t-1}|\mathbf{y}_{1:t-1})$. We begin by computing the gain $\mathbfcal{L}_t^\ell$. After defining $\mathbf{X}^{\ell}_{t}=\mathbf{Q} + \mathbf{A}\bm{\Sigma}_{t-1}^\ell \mathbf{A}^\top$ as in \eqref{xellt},
\begin{align}
	\mathbf{L}_t^\ell &= \mathbf{DG}^\top\mathbf{C}^{\top}(\mathbf{R}\hspace{-0.05cm}+\hspace{-0.05cm}\mathbf{C}\mathbf{X}^{\ell}_{t}\mathbf{C}^{\top}+\mathbf{CG}\mathbf{DG}^\top\mathbf{C}^{\top})^{-1} \notag  \\
	\label{identity1}
	&=\mathbf{DG}^\top\mathbf{C}^{\top}(\mathbf{R}\hspace{-0.05cm}+\hspace{-0.05cm}\mathbf{C}\mathbf{X}^{\ell}_{t}\mathbf{C}^{\top})^{-1}(\mathbf{I}-\mathbf{CG}\mathbf{L}_t^\ell),
\end{align}
which implies
\begin{align}
	\mathbf{L}_t^\ell &= [\mathbf{D}^{-1}+\mathbf{G}^\top \mathbf{C}^\top(\mathbf{R}\hspace{-0.05cm}+\hspace{-0.05cm}\mathbf{C}\mathbf{X}^{\ell}_{t}\mathbf{C}^{\top})^{-1}\mathbf{CG} ]^{-1} \\
	&\times \mathbf{G}^\top \mathbf{C}^\top (\mathbf{R}\hspace{-0.05cm}+\hspace{-0.05cm}\mathbf{C}\mathbf{X}^{\ell}_{t}\mathbf{C}^{\top})^{-1}. \notag 
\end{align}
Letting $\mathbf{D}^{-1}\to \mathbf{0}$ above leads to $\mathbf{L}_t^\ell\to \mathbfcal{L}_t^\ell$ in \eqref{ltell}, where the nonsingularity of $\mathbf{G}^\top \mathbf{C}^\top(\mathbf{R}\hspace{-0.05cm}+\hspace{-0.05cm}\mathbf{C}\mathbf{X}^{\ell}_{t}\mathbf{C}^{\top})^{-1}\mathbf{CG}$ is guaranteed by Assump. \ref{assumption0}. To obtain \eqref{dt1}, we let $\mathbf{D}^{-1}\to \mathbf{0}$ in \eqref{eqn:lemma_filtering_8} and note that the terms associated with $\mathfrak{d}$ in \eqref{eqn:lemma_filtering_8} after replacing \eqref{eqn:lemma_filtering_10} cancel out when the identity $\mathbfcal{L}_t^{\ell}\mathbf{CG}=\mathbf{I}$ is used. Next, we derive \eqref{xtsise}. Using the same identity as in \eqref{identity1},
\begin{align}
	\mathbf{K}_t^\ell \hspace{-0.03cm}&= (\mathbf{X}^{\ell}_{t}\hspace{-0.08cm}+\hspace{-0.05cm}\mathbf{GDG}^{\hspace{-0.03cm}\top}\hspace{-0.02cm})\mathbf{C}^\top\hspace{-0.05cm} (\mathbf{R}\hspace{-0.04cm}+\hspace{-0.05cm}\mathbf{C}\mathbf{X}^{\ell}_{t}\hspace{-0.02cm}\mathbf{C}^{\hspace{-0.02cm}\top}\hspace{-0.12cm} +\hspace{-0.05cm} \mathbf{CGDG}^{\hspace{-0.05cm}\top}\hspace{-0.04cm} \mathbf{C}^{\hspace{-0.03cm}\top}\hspace{-0.02cm})^{\hspace{-0.03cm}-1} \notag \\
	&= \mathbf{G}\mathbf{L}_t^\ell+\mathbf{X}^{\ell}_{t}\mathbf{C}^\top (\mathbf{R}+\mathbf{C}\mathbf{X}^{\ell}_{t}\mathbf{C}^\top)^{-1}(\mathbf{I}-\mathbf{CG}\mathbf{L}_t^\ell) \notag \\
	&\xrightarrow{\mathbf{D}^{-1} \to \mathbf{0}} \mathbf{G}\mathbfcal{L}_t^\ell+\mathbfcal{J}_t^\ell(\mathbf{I}-\mathbf{CG}\mathbfcal{L}_t^\ell), \notag 
\end{align}
where $\mathbfcal{J}_t^\ell$ is defined in \eqref{jtell}. Thus, \eqref{eqn:lemma_filtering_4} combined with \eqref{eqn:lemma_filtering_9} give $\hat{\mathbf{x}}_{t|t}^{k}\to \hat{\mathbf{x}}_{t}^{k}$ as $\mathbf{D}^{-1}\to \mathbf{0}$, where
\begin{align}
	\hat{\mathbf{x}}_{t}^{k}\hspace{-0.08cm}&=\mathbf{A}\hat{\mathbf{x}}_{t-1}^{\ell}+\mathbf{G}\mathfrak{d}+[\mathbf{G}\mathbfcal{L}_t^\ell+\mathbfcal{J}_t^\ell(\mathbf{I}-\mathbf{CG}\mathbfcal{L}_t^\ell)] \notag \\
	&\hspace{0.4cm}\times (\zeta_t^{\kappa}(\mathbf{y}_t)-\bm{\mu}_t^\kappa-\mathbf{C}\mathbf{A}\hat{\mathbf{x}}_{t-1}^{\ell}-\mathbf{C}\mathbf{G}\mathfrak{d}) \notag \\
	&=\hspace{-0.04cm}\mathbf{A}\hat{\mathbf{x}}_{t\hspace{-0.02cm}-\hspace{-0.025cm}1}^{\ell}\hspace{-0.1cm}+\hspace{-0.08cm}[\mathbf{G}\hspace{-0.01cm}\mathbfcal{L}_t^\ell\hspace{-0.08cm}+\hspace{-0.09cm}\mathbfcal{J}_t^\ell\hspace{-0.02cm}(\mathbf{I}\hspace{-0.06cm}-\hspace{-0.06cm}\mathbf{CG}\hspace{-0.02cm}\mathbfcal{L}_t^\ell)](\hspace{-0.01cm}\zeta_t^{\kappa}\hspace{-0.02cm}(\mathbf{y}_t\hspace{-0.01cm})\hspace{-0.07cm}-\hspace{-0.07cm}\bm{\mu}_t^\kappa\hspace{-0.11cm}-\hspace{-0.07cm}\mathbf{C}\hspace{-0.02cm}\mathbf{A}\hat{\mathbf{x}}_{t\hspace{-0.02cm}-\hspace{-0.02cm}1}^{\ell}\hspace{-0.01cm}) \notag \\
	&=\hspace{-0.04cm}\mathbf{A}\hat{\mathbf{x}}_{t\hspace{-0.01cm}-\hspace{-0.01cm}1}^{\ell}\hspace{-0.11cm}+\hspace{-0.07cm}\mathbf{G}\hat{\mathbf{d}}_{t\hspace{-0.02cm}-\hspace{-0.02cm}1}^{k}\hspace{-0.12cm}+\hspace{-0.08cm}\mathbfcal{J}_t^\ell \hspace{-0.03cm}(\hspace{-0.01cm}\zeta_t^{\kappa}\hspace{-0.02cm}(\mathbf{y}_{\hspace{-0.02cm}t}\hspace{-0.01cm})\hspace{-0.09cm}-\hspace{-0.1cm}\bm{\mu}_t^{\hspace{-0.02cm}\kappa}\hspace{-0.11cm}-\hspace{-0.09cm}\mathbf{C}\mathbf{A}\hat{\mathbf{x}}_{t\hspace{-0.01cm}-\hspace{-0.01cm}1}^{\ell}\hspace{-0.12cm}-\hspace{-0.09cm}\mathbf{CG}\hat{\mathbf{d}}_{t\hspace{-0.01cm}-\hspace{-0.01cm}1}^{k}\hspace{-0.02cm}), \notag 
\end{align}
where the terms involving $\mathfrak{d}$ cancel out in the second equality because $\mathbf{K}_t^{\ell}\mathbf{C}\mathbf{G} - \mathbf{G} \to \mathbf{0}$ as $\mathbf{D}^{-1} \to \mathbf{0}$, and where the final equality follows from \eqref{dt1}. Next, we proceed to compute the covariances of each Gaussian component. If we define $\mathbfcal{T}_{t-1}^\ell := \mathbf{G}^\top \mathbf{C}^\top(\mathbf{R}\hspace{-0.05cm}+\hspace{-0.05cm}\mathbf{C}\mathbf{X}^{\ell}_{t}\mathbf{C}^{\top})^{-1}\mathbf{CG}$, then
\begin{align}
	\bm{\Gamma}_{t\hspace{-0.02cm}-\hspace{-0.02cm}1|t}^{k}&\!=\!(\mathbf{I}-\mathbf{L}_t^{\ell}\mathbf{CG})\mathbf{D}\!=\!(\mathbf{I}-[\mathbf{D}^{-1}\hspace{-0.05cm}+\hspace{-0.05cm}\mathbfcal{T}_{t-1}^\ell]^{-1}\mathbfcal{T}_{t-1}^\ell)\mathbf{D} \notag \\
	&\xrightarrow{\mathbf{D}^{-\hspace{-0.02cm}1} \hspace{-0.06cm}\to \mathbf{0}}\hspace{-0.03cm} (\mathbfcal{T}_{t\hspace{-0.02cm}-\hspace{-0.02cm}1}^\ell)^{-1}\hspace{-0.08cm}=:\hspace{-0.06cm}\bm{\Gamma}_{\hspace{-0.02cm}t\hspace{-0.01cm}-\hspace{-0.01cm}1}^{k}\hspace{-0.03cm}. \label{leadingto}
\end{align}
To derive \eqref{sigmat}, we first obtain the intermediate expression
\begin{equation}
	\mathbf{K}_t^\ell \mathbf{CG}\hspace{-0.08cm} = \hspace{-0.08cm}\mathbf{G}\hspace{-0.01cm}(\mathbf{D}^{\hspace{-0.02cm}-\hspace{-0.02cm}1}\hspace{-0.12cm}+\hspace{-0.08cm}\mathbfcal{T}_{\hspace{-0.02cm}t\hspace{-0.02cm}-\hspace{-0.02cm}1}^\ell\hspace{-0.01cm})^{\hspace{-0.03cm}-\hspace{-0.02cm}1}\hspace{-0.01cm}\mathbfcal{T}_{\hspace{-0.02cm}t\hspace{-0.02cm}-\hspace{-0.02cm}1}^\ell\hspace{-0.09cm}+\hspace{-0.06cm} \mathbfcal{J}_t^\ell\mathbf{CG}(\mathbf{D}^{-\hspace{-0.02cm}1}\hspace{-0.12cm}+\hspace{-0.08cm}\mathbfcal{T}_{\hspace{-0.02cm}t\hspace{-0.02cm}-\hspace{-0.02cm}1}^\ell)^{\hspace{-0.03cm}-\hspace{-0.02cm}1}\mathbf{D}^{\hspace{-0.02cm}-\hspace{-0.02cm}1}\hspace{-0.05cm}, \notag 
\end{equation}
which leads to
\begin{align}
	\mathbf{GD}-\mathbf{K}_t^{\ell}\mathbf{CGD} =& (\mathbf{I}-\mathbfcal{J}_t^\ell\mathbf{C})\mathbf{G}(\mathbf{D}^{-1}+\mathbfcal{T}_{t-1}^\ell)^{-1} \notag \\
	\xrightarrow{\mathbf{D}^{-1} \to \mathbf{0}}& (\mathbf{I}-\mathbfcal{J}_t^\ell\mathbf{C})\mathbf{G}\bm{\Gamma}_{t-1}^{k}. \notag
\end{align}
Using this fact, we conclude that $\bm{\Sigma}_{t|t}^{k}\xrightarrow{\mathbf{D}^{-1} \to \mathbf{0}}\bm{\Sigma}_{t}^{k}$, where
\begin{align}
	\bm{\Sigma}_{t}^{k}&=(\mathbf{I}-\mathbfcal{J}_t^{\ell}\mathbf{C})\mathbf{G}\bm{\Gamma}_{t-1}^k\mathbf{G}^{\top} \notag \\
	&\hspace{0.3cm}+(\mathbf{I}-\mathbf{G}\mathbfcal{L}_t^{\ell}\mathbf{C}-\mathbfcal{J}_t^{\ell}\mathbf{C}+\mathbfcal{J}_t^{\ell}\mathbf{CG}\mathbfcal{L}_t^{\ell}\mathbf{C})\mathbf{X}^{\ell}_{t}  \notag \\
	&=(\mathbf{I}-\mathbfcal{J}_t^{\ell}\mathbf{C})(\mathbf{X}^{\ell}_{t}- \mathbf{G}\mathbfcal{L}^{\ell}_{t}\mathbf{C}\mathbf{X}^{\ell}_{t}+\mathbf{G}\bm{\Gamma}_{t-1}^{k}\mathbf{G}^{\top}). \notag
\end{align}
The final result concerns the weights $\delta_t^k$. These quantities, for finite $\mathbf{D}$, are given by 
\begin{equation}
	\label{rearranging}
	\delta_{t|t}^{k} = \frac{\varphi_t^{\kappa} \delta_{t-1|t-1}^{\ell}}{\sum_{\kappa_1=1}^{K}\sum_{\ell_1=1}^{M_{t-1}}\varphi_t^{\kappa_1} \delta_{t-1|t-1}^{\ell_1} \rho_{\kappa,\kappa_1}^{\ell,\ell_1}},   
\end{equation}
where $\rho_{\kappa,\kappa_1}^{\ell,\ell_1}$ is given as the quotient of Gaussian PDFs, which can be expressed as
\begin{align}
	\rho_{\kappa,\kappa_1}^{\ell,\ell_1}\hspace{-0.15cm}&=\hspace{-0.08cm} \sqrt{\hspace{-0.05cm}\frac{\det(\mathbf{R}\hspace{-0.05cm}+\hspace{-0.06cm}\mathbf{C}\mathbf{X}^{\ell}_{t}\mathbf{C}^{\top}\hspace{-0.02cm})\hspace{-0.02cm}\det\hspace{-0.01cm}(\hspace{-0.02cm}\mathbf{I}\hspace{-0.06cm}-\hspace{-0.05cm}\mathbf{CG}\mathbf{L}_t^{\hspace{-0.02cm}\ell_1}\hspace{-0.05cm})}{\det\hspace{-0.01cm}(\hspace{-0.02cm}\mathbf{R}\hspace{-0.05cm}+\hspace{-0.07cm}\mathbf{C}\mathbf{X}^{\ell_1}_{t}\hspace{-0.02cm}\mathbf{C}^{\top}\hspace{-0.05cm})\hspace{-0.02cm}\det(\mathbf{I}\hspace{-0.05cm}-\hspace{-0.05cm}\mathbf{CG}\mathbf{L}_t^\ell)}}\hspace{-0.01cm}\exp\hspace{-0.1cm}\left(\hspace{-0.09cm}-\hspace{-0.02cm}\frac{1}{2}z_{\kappa,\kappa_1}^{\ell,\ell_1}\hspace{-0.05cm}\right)\hspace{-0.07cm}, \notag 
\end{align}
where the identity used in \eqref{identity1} has been applied, and where $z_{\kappa,\kappa_1}^{\ell,\ell_1}\to 0$ as $\mathbf{D}^{-1}\to \mathbf{0}$. By the determinant property in \cite[1.3.P28]{horn2012} and the derivation leading to \eqref{leadingto}, we obtain
\begin{align}
	\rho_{\kappa,\kappa_1}^{\ell,\ell_1} \hspace{-0.09cm}&=\hspace{-0.08cm} \sqrt{\hspace{-0.04cm}\frac{\det\hspace{-0.02cm}(\hspace{-0.01cm}\mathbf{R}\hspace{-0.05cm}+\hspace{-0.07cm}\mathbf{C}\mathbf{X}^{\ell}_{t}\hspace{-0.02cm}\mathbf{C}^{\top}\hspace{-0.02cm})\hspace{-0.02cm}\det\hspace{-0.01cm}(\mathbf{D}^{-\hspace{-0.02cm}1}\hspace{-0.08cm}+\hspace{-0.08cm}\mathbfcal{T}_{t\hspace{-0.02cm}-\hspace{-0.02cm}1}^{\ell}\hspace{-0.01cm})}{\det\hspace{-0.02cm}(\hspace{-0.01cm}\mathbf{R}\hspace{-0.05cm}+\hspace{-0.07cm}\mathbf{C}\mathbf{X}^{\ell_1}_{t}\hspace{-0.02cm}\mathbf{C}^{\top}\hspace{-0.02cm})\hspace{-0.02cm}\det\hspace{-0.02cm}(\hspace{-0.02cm}\mathbf{D}^{-\hspace{-0.02cm}1}\hspace{-0.08cm}+\hspace{-0.07cm}\mathbfcal{T}_{t\hspace{-0.02cm}-\hspace{-0.02cm}1}^{\ell_1})}}\hspace{-0.01cm}\exp\hspace{-0.09cm}\left(\hspace{-0.1cm}-\hspace{-0.02cm}\frac{z_{\kappa,\kappa_1}^{\ell,\ell_1}}{2}\hspace{-0.1cm}\right) \notag \\
	\label{using}
	&\xrightarrow{\mathbf{D}^{-1} \to \mathbf{0}} \sqrt{\frac{\det(\mathbf{R}\hspace{-0.05cm}+\hspace{-0.07cm}\mathbf{C}\mathbf{X}^{\ell}_{t}\mathbf{C}^{\top}\hspace{-0.02cm})\hspace{-0.02cm}\det(\mathbfcal{T}_{t-1}^{\ell})}{\det(\mathbf{R}\hspace{-0.05cm}+\hspace{-0.05cm}\mathbf{C}\mathbf{X}^{\ell_1}_{t}\mathbf{C}^{\top}\hspace{-0.02cm})\hspace{-0.02cm}\det(\mathbfcal{T}_{t-1}^{\ell_1})}}. 
\end{align}
Rearranging \eqref{rearranging} using \eqref{using} leads to \eqref{delta1} and \eqref{delta2} as $\mathbf{D}^{-1}\to \mathbf{0}$, which concludes the proof. \hfill $\square$

\vspace{-0.2cm}

\bibliographystyle{IEEEtran}
\bibliography{references}  

\begin{thebibliography}{10}
\providecommand{\url}[1]{#1}
\csname url@samestyle\endcsname
\providecommand{\newblock}{\relax}
\providecommand{\bibinfo}[2]{#2}
\providecommand{\BIBentrySTDinterwordspacing}{\spaceskip=0pt\relax}
\providecommand{\BIBentryALTinterwordstretchfactor}{4}
\providecommand{\BIBentryALTinterwordspacing}{\spaceskip=\fontdimen2\font plus
\BIBentryALTinterwordstretchfactor\fontdimen3\font minus
  \fontdimen4\font\relax}
\providecommand{\BIBforeignlanguage}[2]{{%
\expandafter\ifx\csname l@#1\endcsname\relax
\typeout{** WARNING: IEEEtran.bst: No hyphenation pattern has been}%
\typeout{** loaded for the language `#1'. Using the pattern for}%
\typeout{** the default language instead.}%
\else
\language=\csname l@#1\endcsname
\fi
#2}}
\providecommand{\BIBdecl}{\relax}
\BIBdecl

\bibitem{Hieu2011}
H.~Trinh and T.~Fernando, \emph{Functional Observers for Dynamical
  Systems}.\hskip 1em plus 0.5em minus 0.4em\relax Springer, 2011.

\bibitem{Yao2022}
X.~Yao, V.~Le, and I.~Lee, ``{Unknown Input Observer-Based Series DC Arc Fault
  Detection in DC Microgrids},'' \emph{IEEE Transactions on Power Electronics},
  vol.~37, no.~4, pp. 4708--4718, 2022.

\bibitem{Valikhani2019}
M.~Valikhani and D.~Younesian, ``{Bayesian framework for simultaneous
  input/state estimation in structural and mechanical systems},''
  \emph{Structural Control and Health Monitoring}, vol.~26, pp. 1--25, 2019.

\bibitem{glover1969linear}
J.~Glover, ``The linear estimation of completely unknown signals,'' \emph{IEEE
  Trans. on Automatic Control}, vol.~14, no.~6, pp. 766--767, 1969.

\bibitem{kitanidis1987unbiased}
P.~K. Kitanidis, ``Unbiased minimum-variance linear state estimation,''
  \emph{Automatica}, vol.~23, no.~6, pp. 775--778, 1987.

\bibitem{gillijns2007unbiased}
S.~Gillijns and B.~De~Moor, ``Unbiased minimum-variance input and state
  estimation for linear discrete-time systems,'' \emph{Automatica}, vol.~43,
  no.~1, pp. 111--116, 2007.

\bibitem{hsieh2009extension}
C.-S. Hsieh, ``Extension of unbiased minimum-variance input and state
  estimation for systems with unknown inputs,'' \emph{Automatica}, vol.~45,
  no.~9, pp. 2149--2153, 2009.

\bibitem{yin2025simultaneous}
L.~Yin, W.~Xie, S.~Wang, and V.~Sreeram, ``Simultaneous input and state
  estimation: From a unified least-squares perspective,'' \emph{Automatica},
  vol. 171, \textnormal{Article} 111906, 2025.

\bibitem{bitmead2019kalman}
R.~R. Bitmead, M.~Hovd, and M.~A. Abooshahab, ``{A Kalman-filtering derivation
  of simultaneous input and state estimation},'' \emph{Automatica}, vol. 108,
  \textnormal{Article} 108478, 2019.

\bibitem{Song2024}
X.~Song and W.~X. Zheng, ``{A Kalman-filtering derivation of input and state
  estimation for linear discrete-time systems with direct feedthrough},''
  \emph{Automatica}, vol. 161, \textnormal{Article} 111453, 2024.

\bibitem{gakis2024limit}
G.~Gakis and M.~C. Smith, ``{A limit Kalman filter and smoother for systems
  with unknown inputs},'' \emph{International Journal of Control}, vol.~97,
  no.~3, pp. 532--542, 2024.

\bibitem{Fang2013}
H.~Fang, R.~A. de~Callafon, and J.~Cort{\'{e}}s, ``{Simultaneous input and
  state estimation for nonlinear systems with applications to flow field
  estimation},'' \emph{Automatica}, vol.~49, no.~9, pp. 2805--2812, 2013.

\bibitem{cedeno2021}
A.~L. Cede{\~{n}}o, R.~Albornoz, R.~Carvajal, B.~I. Godoy, and J.~C.
  Ag{\"{u}}ero, ``{On Filtering Methods for State-Space Systems having Binary
  Output Measurements},'' \emph{IFAC-PapersOnLine}, pp. 815--820, 2021.

\bibitem{Cedeno2021b}
------, ``{A Two-Filter Approach for State Estimation Utilizing Quantized
  Output Data},'' \emph{Sensors}, vol.~21, no.~22, p. 7675, 2021.

\bibitem{kim2020simultaneous}
H.~Kim, P.~Guo, M.~Zhu, and P.~Liu, ``Simultaneous input and state estimation
  for stochastic nonlinear systems with additive unknown inputs,''
  \emph{Automatica}, vol. 111, \text{Article} 108588, 2020.

\bibitem{tian2024composite}
B.~Tian, W.~Li, X.~Yu, W.~Wang, and L.~Guo, ``{Composite disturbances nonlinear
  filtering for simultaneous state and unknown input estimation under
  non-Gaussian noises},'' \emph{IEEE Transactions on Instrumentation and
  Measurement}, vol.~73, pp. 1--10, 2024.

\bibitem{Zou2020}
L.~Zou, Z.~Wang, J.~Hu, and D.~Zhou, ``{Moving horizon estimation with unknown
  inputs under dynamic quantization effects},'' \emph{IEEE Trans. on Automatic
  Control}, vol.~65, no.~12, pp. 5368--5375, 2020.

\bibitem{yong2016unified}
S.~Z. Yong, M.~Zhu, and E.~Frazzoli, ``A unified filter for simultaneous input
  and state estimation of linear discrete-time stochastic systems,''
  \emph{Automatica}, vol.~63, pp. 321--329, 2016.

\bibitem{sarkka2013bayesian}
S.~S{\"a}rkk{\"a}, \emph{{Bayesian Filtering and Smoothing}}.\hskip 1em plus
  0.5em minus 0.4em\relax Cambridge University Press, 2013.

\bibitem{davis2007methods}
P.~J. Davis and P.~Rabinowitz, \emph{Methods of Numerical Integration}.\hskip
  1em plus 0.5em minus 0.4em\relax Courier Corporation, 2007.

\bibitem{kitagawa1994two}
G.~Kitagawa, ``The two-filter formula for smoothing and an implementation of
  the {G}aussian-sum smoother,'' \emph{Annals of the Institute of Statistical
  Mathematics}, vol.~46, no.~4, pp. 605--623, 1994.

\bibitem{runnalls2007kullback}
A.~R. Runnalls, ``Kullback-{L}eibler approach to {G}aussian mixture
  reduction,'' \emph{IEEE Transactions on Aerospace and Electronic Systems},
  vol.~43, no.~3, pp. 989--999, 2007.

\bibitem{abooshahab2022simultaneous}
M.~A. Abooshahab, M.~Alyaseen, R.~R. Bitmead, and M.~Hovd, ``Simultaneous input
  \& state estimation, singular filtering and stability,'' \emph{Automatica},
  vol. 137, \textnormal{Article} 110017, 2022.

\bibitem{cedeno2024gaussian}
A.~L. Cede{\~n}o, R.~A. Gonz{\'a}lez, and J.~C. Ag{\"u}ero, ``Gaussian sum
  filtering for {Wiener} state-space models with a class of non-monotonic
  piecewise nonlinearities,'' \emph{IFAC-PapersOnLine}, pp. 25--30, 2024.

\bibitem{petersen2012}
K.~B. Petersen and M.~S. Pedersen, \emph{{The Matrix Cookbook}}.\hskip 1em plus
  0.5em minus 0.4em\relax Technical University of Denmark, 2012.

\bibitem{horn2012}
R.~A. Horn and C.~R. Johnson, \emph{Matrix Analysis, \textnormal{2nd
  Edition}}.\hskip 1em plus 0.5em minus 0.4em\relax Cambridge University Press,
  2012.

\end{thebibliography}

\end{document}